\documentclass[preprint,10pt]{elsarticle}

\usepackage{graphicx}
\usepackage{amssymb}

\usepackage{lineno}
\journal{arXiv}

\begin{document}

\begin{frontmatter}


\title{Financial Market Prediction}



\author{Mike H. Wu}

\address{Undergraduate, Yale University}

\begin{abstract}
Given financial data from popular sites like Yahoo and the London Exchange, the presented paper attempts to model and predict stocks that can be considered "good investments". Stocks are characterized by 125 features ranging from gross domestic product to EDIBTA, and are labeled by discrepancies between stock and market price returns. An artificial neural network (Self-Organizing Map) is fitted to train on more than a million data points to predict "good investments" given testing stocks from 2013 and after. 
\end{abstract}


\end{frontmatter}

\section{Introduction}
\label{S:1}
\subsection{Statement of Purpose}
The motivation for such a project is the design and usage of modern machine learning algorithms in the financial field. Highlighted are the following points of original content:
	\begin{itemize}
    	\item Tuning of a cumulative summation algorithm using exhaustive search and post-consolidation.
        \item Labeling testing data on a trained map using weighted, fractional, and convolved SOM units.  
    \end{itemize}

\subsection{Background Information}
Much of successful financial investment is based in experience or chance, making it hard to extend those same principles to new stocks at arbitrary time intervals. Instead, it would be desirable to design a mechanism to automate the investment process, fully describing what it means to be a "successful stock" and being able to recognize future examples of such.\\

Most examples of financial forecasting algorithms use predefined indicators from manual analysis. These often require pre-specified period length, and are often full of non-linearity and irregularity, producing highly orchestrated and arguably inconsistent models. Similar attempts have previously incorporated various other artificial neural networks like SVM and LSSVM. In this particular case, the SOM is chosen for its ability to decompose heterogeneous data points into several homogeneous regions. Given the statistical similarity in the mentioned homogeneous regions, it is likely that the SOM may be able to capture the non-stationary properties of a financial series. 



\section{Methods}
\label{S:2}

The following sections define an overview of the algorithm employed for market prediction. 

\subsection{Data Extraction}


Financial statistics for a large portion of existing companies worldwide are available for free online. For the scope of this project, all data was extracted from the London Stock Exchange and Yahoo Finance. Due to limitations of availability, only companies in the NYSE and NASDAQ markets were considered (future work may incorporate FTSE/LSE).\\

Several constraints were upheld to ensure the quality of future testing/training, some of which are:
\begin{itemize}
	\item Only companies with at least 9 years of available quarterly reports are considered for inclusion. 
    \item Only companies with adequate fundamental statistical data are considered for inclusion.
\end{itemize}

With the proper constraints, initial data was drawn from over 6,500 companies, each bearing data from 40 quarters (10 years). Each company contained data regarding stock prices, comparable market prices, and a list of 125 feature statistics, including but not limited to PE Ratio, Gross Profit Margin, Interest, Inventory, Intangibles, Debt, Industry Comparison Rates, EBIDTA, Sales, Cash Flow, etc. 

\subsection{Feature Vector Creation}

Given a large amount of data, it is important to format the selection into vectors (i.e. arrays of features). All vectors are stored as Numpy arrays for efficiency and for vectorized operation usage. \\

An important part of creating vectors is "cherry picking" which rows of data to keep. All vectors are organized by start and end time such that any vectors with time intervals where no stock or market price data is available is immediately discarded. 22 out of 2500 such companies were tossed due to missing price data. For market price, the ticker GPSC was used to estimate S\&P500 fluctuations, which were seen as a good representative of market changes. In addition, because ratios are good for categorizing data, vectors without data for important ratios such as PE or EDIBTA were discarded. All other missing data were replaced with Numpy NaN's (because financial data is considered messy, about 30\% of all data were NaN's).\\

Given a feature matrix with NaN's, it's a good notion to calculate the distribution of the missing data. From this project's data set, because only 41 out of 122 features contained more than 1\% of NaN or Inf values, those 41 were discarded, leaving 81 features for consideration. The features removed were confirmed to not be fundamental data, mostly being percentage comparisons to other companies or string values. However, because these removed features averaged 70\% NaN's, there is justification to discarding them. 



\subsection{Feature Selection}

Depending on the model used, the number of features should be different. Although the Self-Organizing Map (SOM) is rather lenient on feature count (i.e. performs a variation of feature selection itself), it is still important to reduce the features to more reasonable number (i.e. $\ll$ than 81).\\

Three different methods were implemented: Principle Component Analysis (PCA), Extra Trees Classifier (EXT), Recursive Feature Extraction (RFE). These three were chosen for their preservation of feature identity. Given any interested user of such an algorithm, it is desirable to return to him/her the features that best explain the variance in the data and the choices made. RFE and EXT merely rank features, and PCA produces a decipherable linear combination of vectors, all easily understood and explained.

\begin{figure}[h]
  \centering\includegraphics[width=0.6\linewidth]{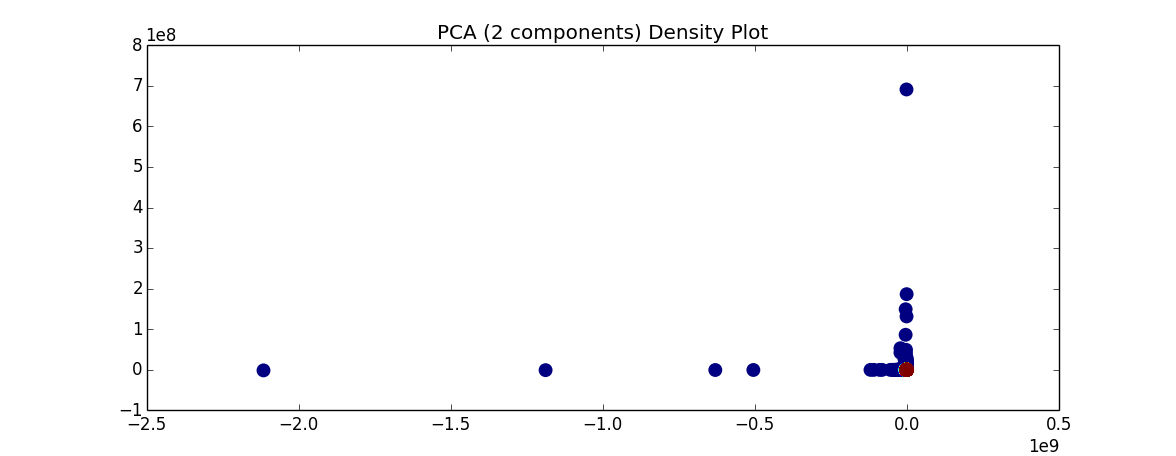}
  \caption{PCA can easily be visualized but by dimensionality reduction, much of the data is indistinguishable. The density is highly concentrated in 1 point (not allowing any future separation).}
\end{figure}

\begin{itemize}
	\item \subsubsection{PCA} PCA is an intuitive choice given that it does not require labels to be effective. It can also easily characterize which features explain exactly what percentage of the variance. Given the vectors in this dataset, 2 principle components explained 97\% of the variance and 7 explained 99.999\%. However, although the percentages are positive signals, plotting the 2 principle components shows that the majority of the data lies in a single dense clump. It will be extremely hard to differentiate good and bad investments by any linear/non-linear classifier given the dimensionality reduction. Increasing the number of principle components did not remedy the situation. 
    \item \subsubsection{RFE} RFE is a simple concept; it picks and replaces features to form different combinations to find the best combination. While it does produce ranked features, given the relevant dataset of 1 million vectors and 122 features, the running time is too expensive. Additionally, classical RFE usage assumes a simple linear classifier (such as SVM with a linear kernel or Logistic Regression, neither of which accurately represent the model intended for use).  
    \item \subsubsection{EXT} EXT is a variation of a random forest feature selection. It is much more efficient than RFE and produces a complete ranked list of features. However, because this algorithm depends on randomness, successive runs may not converge to the same features. 
\end{itemize}

By functionality and efficiency, EXT was chosen as the "best" selection tool for this data set. The top 25 features were kept. This is an arbitrary over-estimation, relying on SOM for the remainder of the feature filtering. 

\subsection{Vector Labeling (CUSUM)} 

Note: The labels for the vectors are determined by the mentioned log stock and log market prices. Refer to the data extraction code for details (appendix).\\

\begin{figure}[h]
  \centering\includegraphics[width=0.6\linewidth]{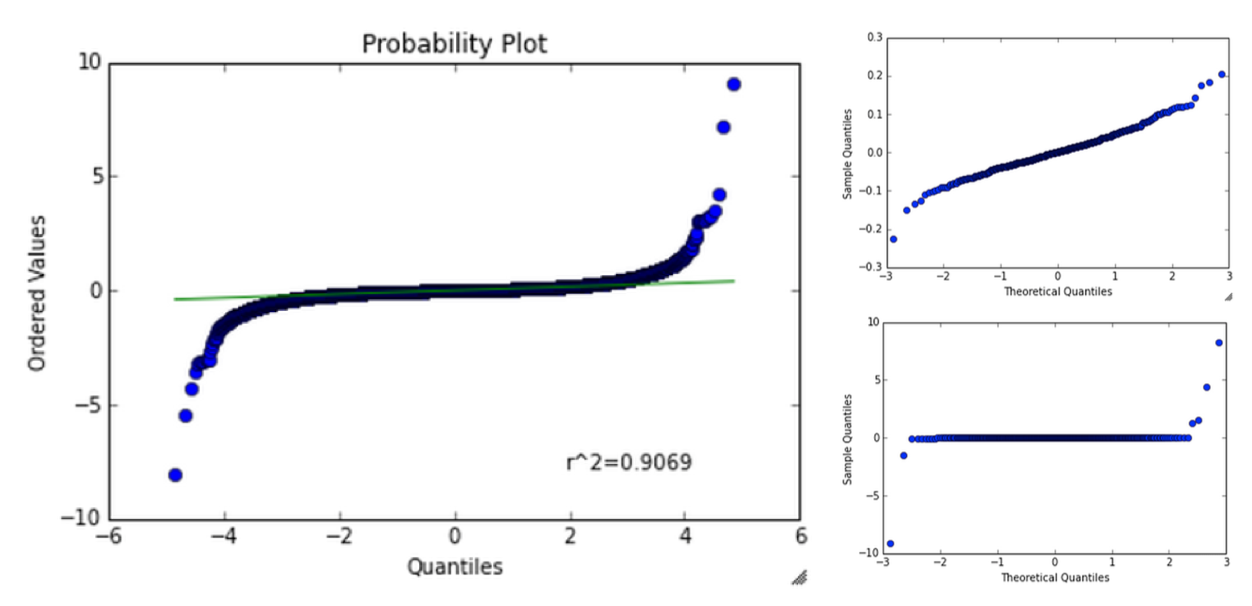}
  \caption{QQ-plots for log stock and log index data. (\textit{left}) Plot for all log stock price data for all tickers. Notice that while a large portion of the data is normal, there are fat tails, suggesting deviation from normality at both ends of the distribution. (\textit{top right}) Log stock data for a single ticker (AA). (\textit{bottom right}) Log market data for a single ticker (AA).}
\end{figure}

Although SOM is an unsupervised classification method, to properly tune its settings, it is helpful to know the labels of the training data. For this project, CUSUM was used to detect time points of rapid shifts in the cumulative sum of the log stock price data. Given some array of detected time points: $t_1$, $t_2$, $t_3$, ..., $t_n$, extract the corresponding vectors: $V_{t_1}$, $V_{t_2}$, $V_{t_3}$, ..., $V_{t_n}$. For each time interval: ($t_1$, $t_2$), ($t_2$, $t_3$), ..., ($t_{n-1}$, $t_n$), one can test for normality. If normal, a two-sample, one-tailed t-test is applied with the hypothesis being that the stock data and the market data are from the same distribution (reject with 95\% confidence). If not normal, a comparable non-parametric t-test (i.e. Mann Whitney) can be employed although the t-test should be relatively lenient with slight non-normality. Based on the hypothesis test, label the vector at the start of the time interval 0 or 1 (w/ 1 being a "good" investment). \\

If not enough data remains post-labeling, it is possible to include a "sliding window": Given time interval ($t_1$, $t_2$), one can use the explained procedure to label $V_{t_1}$. One can also use time interval ($t_{1+j}$, $t_2$) to label $V_{t_{1+j}}$ for $\forall{j}$ s.t. size($t_1$, $t_2$) $>$ some minimum sample size. If not, ignore the interval.\\

\begin{figure}[h]
  \centering\includegraphics[width=0.85\linewidth]{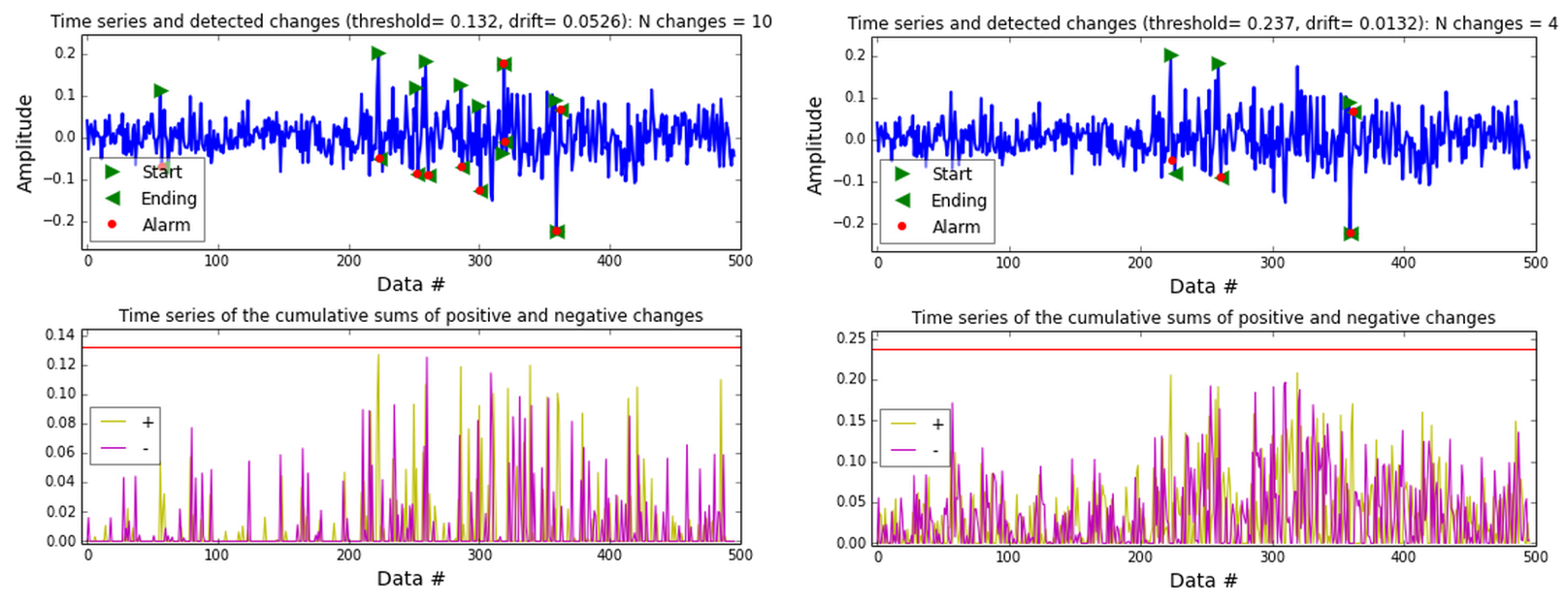}
  \caption{Two examples of the CUSUM algorithm on the same dataset given two "target points" (2 sensitivity levels). The right image is less sensitive, and therefore ignore more cumulative summation changes. Notice that there is not a set interval size. }
\end{figure}

The one caveat is tuning the CUSUM, which is dependent on a threshold and a drift parameter, which both define the function sensitivity level. In order to tune the CUSUM, any given parameters produce an output with segments of extreme sizes, which may prove to be difficult for vector comparisons during model training/testing. Instead, it is possible to give CUSUM an interval size to "aim for" while preserving cumulative summation change detection. Specifically, use exhaustive search (double for loop) to find any set of parameters that is hypersensitive for change (i.e. a large amount of small segments). Then based on the direction of CUSUM points on the graph topology (i.e. heading towards a local maximum or minimum), consolidate CUSUM intervals until close to the "target size". The logic is then to avoid several segments that explain the same upward or downward slope (eliminating unnecessary middle points). This may still result in many intervals of varying sizes but should help reduce unnecessary variation. This tuning process is expensive but reasonable for 3000-sized company data. \\

Given the ability to "control" CUSUM, there are 3 different labeling techniques: small (25 weeks), medium (52 weeks), large (156 weeks). After each labeling, the output is 850,000 vectors with 850,000 corresponding labels, with each vector bearing 25 indicies. 

\subsection{Model Usage (SOM)}
Tuning the SOM is somewhat of an art. Given the number of feature selection techniques, and the number of labeling options, there are several combinations available. \\

\begin{figure}[h]
  \centering\includegraphics[width=0.4\linewidth]{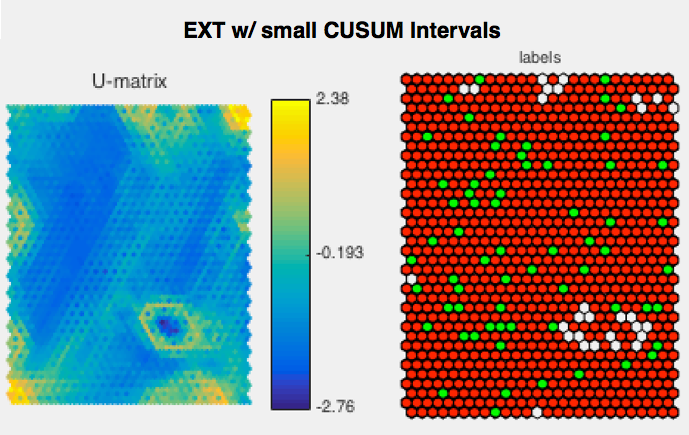}
  \caption{(\textit{left}) U-matrix where blue represents nodes that are close together. (\textit{right}) Projected LCP where green represents a good investment. Notice that there are no clear clusters of "good investments" suggesting randomness.}
\end{figure}

A U-matrix (UMAT) is the matrix of euclidean distances between the codebook vectors of neighboring neurons, and a labeled Component Plane (LCP) is defined as the projection of the known training labels onto a slice of the UMAT. If the good and bad clusters in the LCP correspond to clusters of "close" nodes in the UMAT, then the label can be considered good. 

\subsubsection{Observations}
\begin{itemize}
	\item PCA with any labeling scheme produced poor results. Although the good clusters in the LCP matched with the UMAT, there is a seemingly random mixture of good an bad mini-clusters within it, making it indivisible. 
    \item EXT and RFE produced very similar results. The choice between the two may not be particularly important, especially seeing that the two often prioritize the same features. 
    \item Medium-sized CUSUM is the best choice. Small-sized CUSUM does not seem to define clear clusters in the LCP, while large-sized CUSUM seems to blur everything together, meaning that there are clusters of barely good and barely bad investments. 
\end{itemize}

\begin{figure}[h]
  \centering\includegraphics[width=0.8\linewidth]{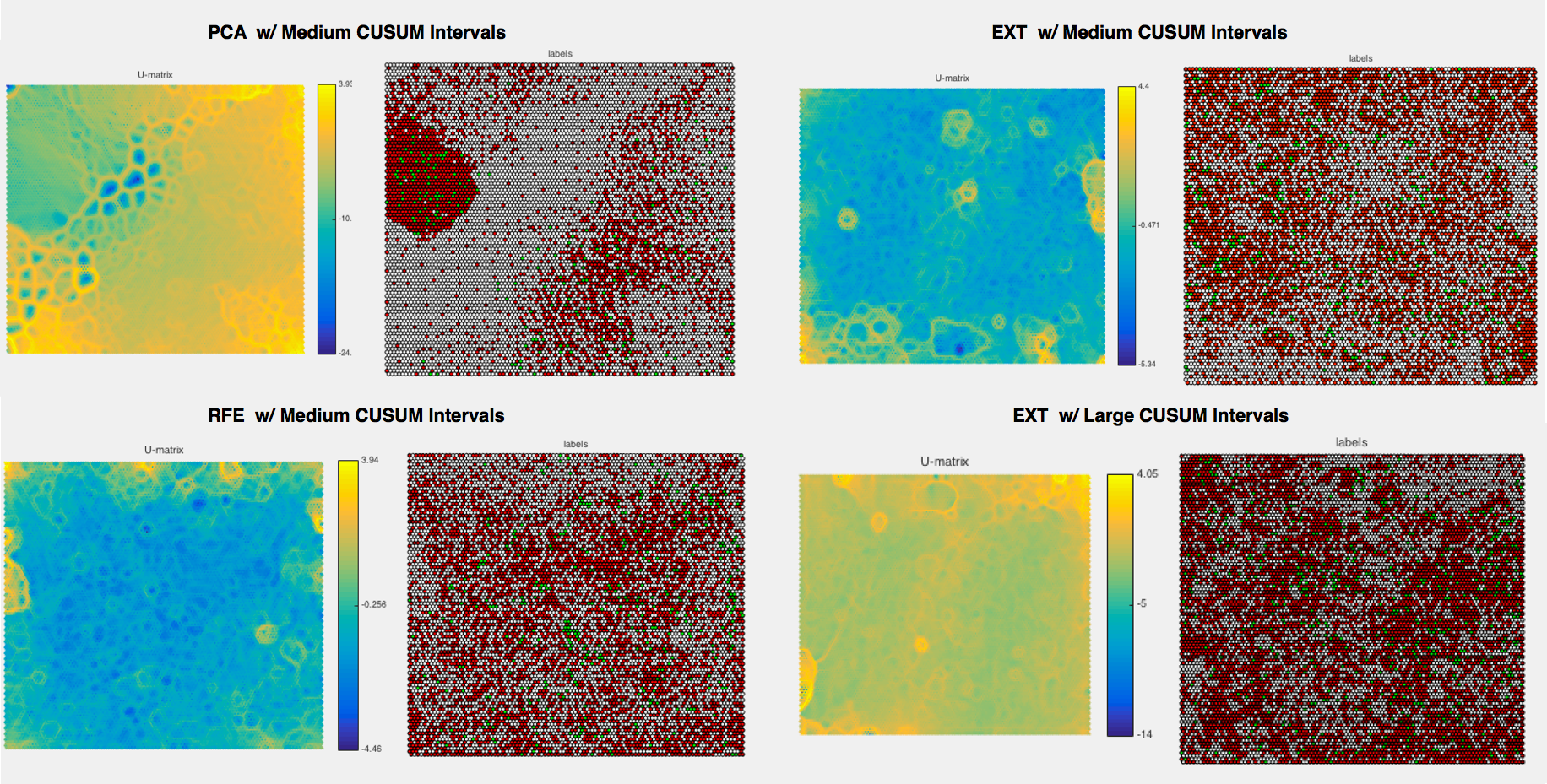}
  \caption{Four different possible combinations for SOM tuning. Notice that RFE and EXT with medium CUSUM intervals are very similar and have arguably similar LCP and UMAT clusters.}
\end{figure}

That provided, the final configuration was an EXT with Medium-sized CUSUM labeling using both parametric and non-parametric t-tests depending on each interval normality. A SOM map of size 50 by 50 is used trained on all 850,000 training vectors. See MATLAB's Somtoolbox for implementation details. 

\subsection{Testing Procedure}
Given a trained model, and the testing data from 2013-Present, one can label the testing vectors with the SOM. General labeling is performed by finding the best matching SOM unit (BMU) for each testing vector given a distance criterion. Generally, because multiple vectors are projected to the same SOM units, each one can vote for that SOM unit to be labeled as a good or bad investment unit. The vector is then labeled by majority vote. 

\subsubsection{Custom Additions}
\begin{itemize}
	\item \textit{Fractionize} \\
    	Instead of taking majority vote; it is more interesting to retain the fraction of good investment votes to total votes for each SOM unit. This allows for ranking and less loss of information.
        
	\item \textit{Weighting} \\
    	Given the nature of fractions, 1/1 is equivalent to 23/23. However, in context, 1 "good" vote out of 1 vote total is much weaker than 23 "good" votes out of 23 total votes. To adjust for this, let $g_{i}$ be the number of "good" votes, $b_{i}$ be the number of "bad" votes, and let $w_{i}$ be a new variable representing a weight. Then for each vector $i$, the label should be edited to $w_{i} * (g_{i} / b_{i})$. The simplest example would be $w_{i} = g_{i}$.
        
    \item \textit{Convolution} \\
    	Given the nature of a matrix, it is more convincing that a SOM unit represents a "good investment" if the neighbors also represent "good investments" rather than a singleton SOM unit. To mathematically represent this, given some gaussian kernel (i.e. size 5), one can convolve the gaussian kernel against the matrix of fractional-weighted SOM units while preserving matrix size. This will then further differentiate units, and produce a desired blurring effect, more precisely ranking units.  
\end{itemize}

Finally, to label any vector, find its BMU and map it to the fractional-weighted-convolved (FWC) matrix. To find the top $N$ companies, rank the vectors by the matrix value and find each one's corresponding ticker. 

\begin{figure}[h]
  \centering\includegraphics[width=0.95\linewidth]{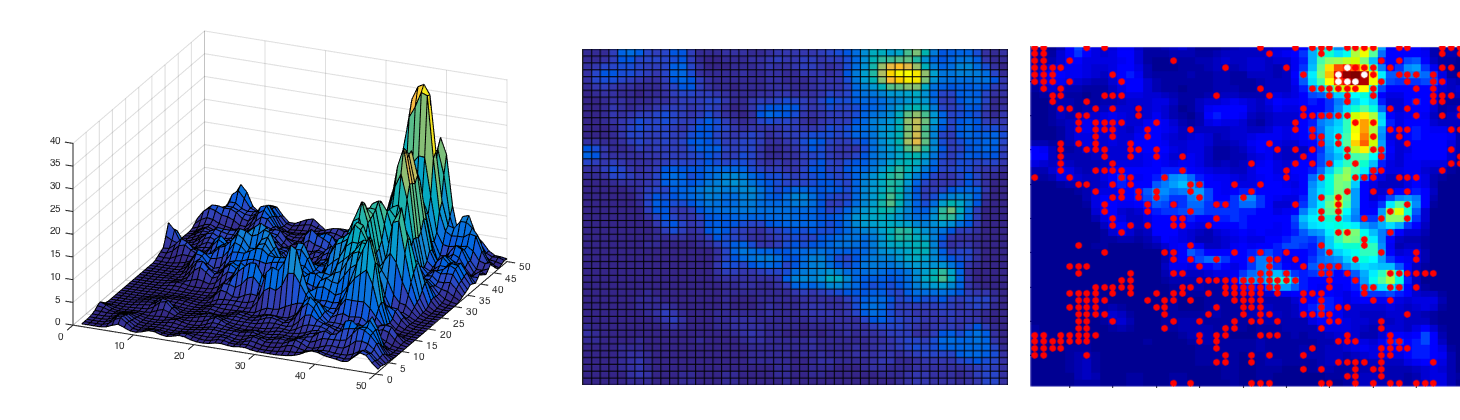}
  \caption{(\textit{left}) Contour plot of FWC. (\textit{middle}) Flat projection of FWC. (\textit{right}) Testing vectors projected onto the FWC. Notice that white points (top ten) lie in the "highest" regions of the FWC. These top vectors are very close to each other in SOM space.}
\end{figure}

\subsection{Additional Algorithms}
\subsubsection{Minimum Sample Size Calculation}
This calculation is very important because an interval too small would lead to wrong estimates and wrong conclusions. As an example, assume that the client for this project decided that the minimum annual relative return acceptable is 5\%, and to ensure 80\% test power for a particular effect size X. \\

Let $\pi(\tau)$ be the power, $n$ be the sample size, $\phi$ be a quantile function of the normal distribution, $\sigma^{2}$ be the sample variance. In addition, let $t_{start}$, $t_{end}$ be the time point at the start, end of some interval. Define stock return $S^{R}$ as $S_{t_{end}} / S_{t_{start}}$ where $S_{x}$ is the price of the stock at time $x$. Define the market return as $M^{R}$ as $M_{t_{end}} / M_{t_{start}}$.\\

\begin{enumerate}
	\item $\pi(\tau) \approx 1 - \phi(1.64 - (\tau * \sqrt{n}) / \sigma) > 0.80$
    \item $ 0.20 > \phi(1.64 - (\tau * \sqrt{n}) / \sigma)$
    \item $\tau*\sqrt{n} / \sigma > 1.64 - zscore(0.20) = 1.64 + 0.84 = 2.48 $
    \item $\sqrt{n} > 2.48*\sigma / \tau$
    \item $n > (2.48*\sigma / \tau)^{2}$
    \item $S^{R} = 0.05 + M^{R}$
    \item $\tau = \log{(0.05+M^{R}) / M^{R}}$ if $S^{R} \equiv M^{R}$
    
\end{enumerate}

For each interval in CUSUM, if the number of samples in that interval is less than $n$, skip the interval. 

\section{Results}
\label{S:3}

\subsection{Features Selected}
The following are the features chosen to be used for training/testing. \\

High PE Ratio, long-term debt/total capital, pre-tax profit margin, EBITDA Per Share, Return on Assets (ROA), Normalized Close PE Ratio, Shares Outstanding, cash flow per share, price/tangible book ratio, post-tax profit margin, working capital as \% of price, price/equity ratio, price/revenue ratio, interest as \% of invested capital, normalized net profit margin, total assets per share, R\&D as \% of Revenue, Normalized ROA, net profit margin, working capital per share, Dividend Per Share, Return on Average Equity, Normalized ROE, leverage ratio, sales per \$ inventory.

\subsection{Top Ten Companies To Invest In}
The following are the raw tickers depicting the top 10 companies (in order) to invest in (given testing data in 2013 and beyond).\\

MICT, CXDC, ITRN, TDG, XIN, WX, FTK, SPU, GURE, PWRD

\section{Discussion/Rationale}
\label{S:4}

\subsection{Features Selected}
It is interesting to note that many of features chosen overlap with features mentioned in literature, providing some informal credibility. For example, Browne's novel \textit{The Little Book of Value Investing} speaks to a similar ranking, emphasizing awareness for price-earnings (PE) ratios, debt ratios, profit margins, returns, and price-to-book ratios, all of which are ranked very highly of importance by EXT.

\subsection{Top Ten Companies}
As an analysis of the top ten companies chosen, it is educational to do a short description of each company.

\begin{enumerate}
	\item Ticker: MICT \\
  		Name: Micronet Enertec Technologies, Inc. \\
        Dates: April 1st, 2013 to June 30th, 2013 \\ 
        Industry: Aerospace/Defense (Industrial Goods)\\
        Location: New Jersey, US
    \item Ticker: CXDC \\
    	Name: China XD Plastics Company Ltd. \\
        Dates: April 1st, 2013 to June 30th, 2013 \\ 
        Industry: Rubber \& Plastics (Consumer Goods)\\
        Location: Harbin, China
    \item Ticker: ITRN \\
    	Name: Ituran Location \& Control Ltd. \\
        Dates: April 1st, 2013 to June 30th, 2013 \\
        Industry: Wireless Communication, GPS, Location Technology (Technology)\\
        Location: Azour, Israel
    \item Ticker: TDG \\ 
    	Name: TransDigm Group Incorporated \\
        Dates: July 1st, 2013 to  September 30th, 2013 \\
        Industry: Aerospace/Defense (Industrial Goods) \\ 
        Location: Cleveland, US
    \item Ticker: XIN \\
    	Name: Xinyuan Real Estate Co., Ltd. \\
        Dates: April 1st, 2013 to June 30th, 2013 \\
        Industry: Real Estate Development (Financial) \\ 
        Location: Beijing, China
    \item Ticker: WX \\ 
    	Name: WuXi PharmaTech (Cayman) Inc. \\
        Dates: April 1st, 2013 to June 30th, 2013 \\
        Industry: Medical Laboratories \& Research (Healthcare) \\ 
        Location: Shanghai, China
    \item Ticker: FTK \\
    	Name: Flotek Industries Inc. \\
        Dates: April 1st, 2013 to June 30th, 2013 \\ 
        Industry: Oil \& Gas Equipment \& Services (Basic Materials)\\
        Location: Texas, US
    \item Ticker: SPU \\
    	Name: SkyPeople Fruit Juice, Inc. \\
        Dates: April 1st, 2013 to June 30th, 2013 \\
        Industry: Beverages - Soft Drinks (Consumer Goods) \\ 
        Location: Xi, China
    \item Ticker: GURE \\ 
    	Name: Gulf Resources, Inc. \\
        Dates: April 1st, 2013 to June 30th, 2013
        Industry: Specialty Chemicals (Basic Materials) \\ 
        Location: Shouguang, China
    \item Ticker: PWRD \\ 
    	Name: Perfect World Co., Ltd. \\
        Dates: April 1st, 2013 to June 30th, 2013
        Industry: Multimedia \& Graphics Software (Technology) \\ 
        Location: Beijing, China
\end{enumerate}

The question is "Do these companies make sense as probable companies to be good investments?" Considering that these are from 2013, the answer is likely "yes". Notice primarily that the companies are either in military, consumer goods, oil, or technology, all of which are large fields with large potential for growth. The consumer goods companies are all based in China. In fact, 6 out of the 10 companies are based in China. While surprising, this does make sense since it is the leader in consumer goods production and is in a state of rapid economic growth. The remaining majority are companies based in US specializing in defense technology, which is also not surprising seeing that the US is a superpower with the largest military in the world. In summary, these results are not unbelievable. Politically and economically, there exist good cases to justify each of companies in that list. Furthermore, from a more scientific angle, the top ten companies originated from the same cluster and are extremely close in SOM space, suggesting proper model training.

\section{Conclusion}
\label{S:5}
The question here is "How have these companies been doing since April of 2013?" Although this cannot fully reject or support the results above, it may give some indication to their accuracy. \\

\begin{figure}[h]
\centering\includegraphics[width=0.7\linewidth]{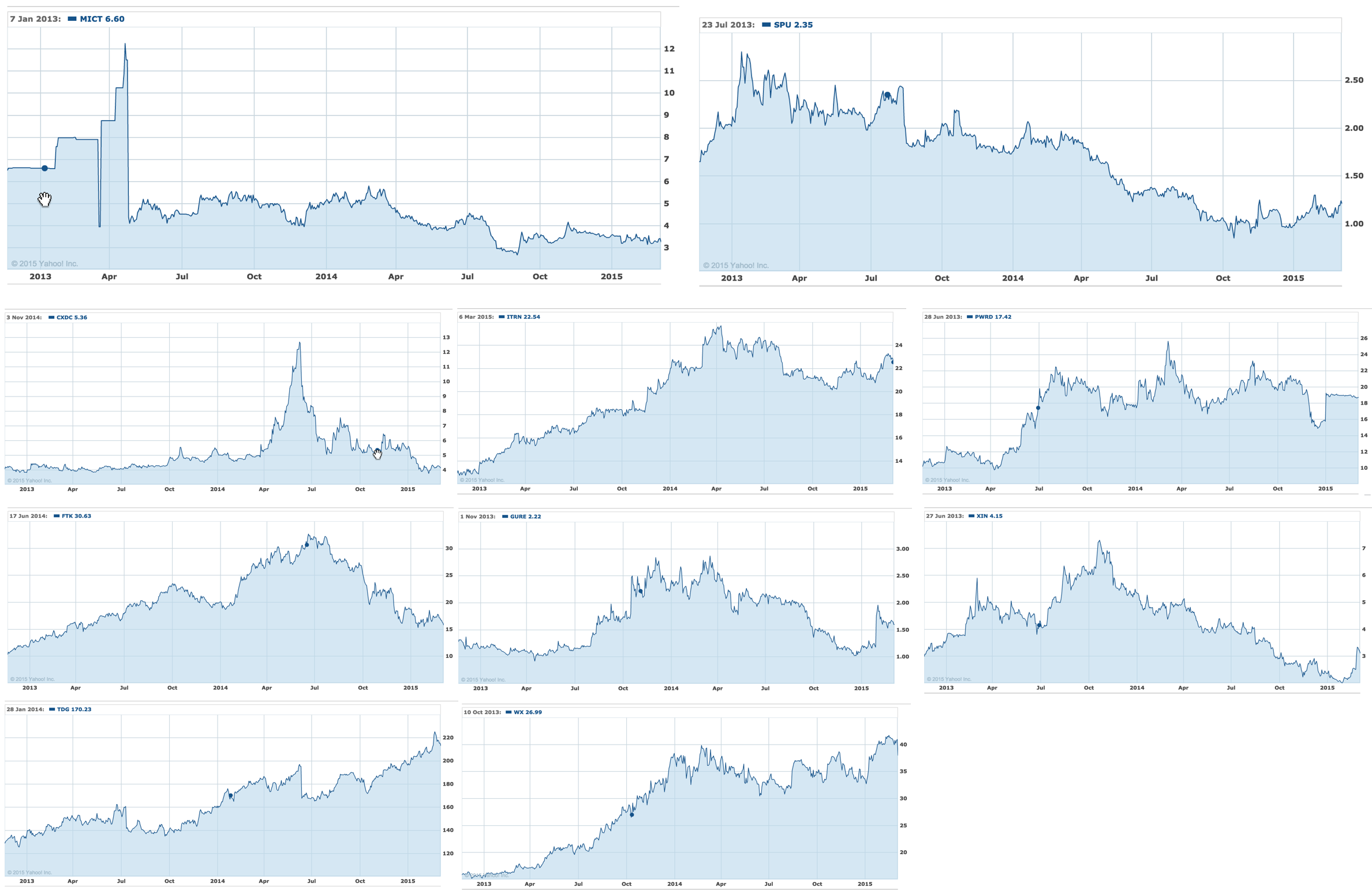}
\caption{Time data of stock prices from April 2013 to January 2015 for each of the top ten stocks. Notice that the bottom 8 exhibit traits of a good investment, notably parabolic shape. The top two:  MICT graph (\textit{top left}), SPU graph (\textit{top right}). }
\end{figure}

With the exception of SPU and MICT, all of the companies chosen as "good investments" had a significant increase in stock prices from April 2013 to January 2015. All of these stock can be described as parabolic, which contains some large peak in between 2013 and 2015. Some of the companies showed large sudden jumps and some others displayed more continuous gradual growth. Note that this model does not provide any assistance on when to sell stocks. However, had a client bought all 10 stocks April 2013, he/she would have gained a large sum of earnings. \\

The exceptions are SPU and MICT. Notice with the MICT graph that there was a particularly large peak in earnings around march that continued to grow after April, hence justifying its position as top 10, but immediately after May of 2013, the stock plummeted and converged to a price below buying price. Therefore, it isn't considered a success. With SPU, the trend showed a slight increase in prices around April but continuous decrease afterwards. \\

In summary, although further testing is needed, the model is successful with the caveat that it may group truly successful stocks with stocks with momentary success (i.e. SPU  or MICT). 


\section*{Acknowledgements}
The author acknowledges the helpful support to this exercise from Irina Higgins.

\section*{Appendix}
All development and production code will be included in the dropbox folder. Please see the following link: https://www.dropbox.com/home/Mike/wu-market-predictor-project. This includes three segments: Python Scripts (production code for feature selection, feature vector creation, results analysis, vector labeling), iPython Notebooks (equivalent development code), and MATLAB scripts (SOM implementation and data scraping).






\end{document}